\documentclass{mem}
\usepackage{natbib}\usepackage{txfonts}\usepackage{balance}
\usepackage{graphicx}
\usepackage[a4paper]{hyperref}

\def\kpc{\mathrm {kpc}}
\def\pc{\mathrm {pc}}
\def\kms{\mathrm {km s}^{-1}}
\def\kmskpc{\mathrm {km s}^{-1}\mathrm{kpc}^{-1}}

\begin{document}
\title{The Milky Way Spiral Arm Pattern}
\subtitle{3D Distribution of Molecular Gas}

\author{
P. \,Englmaier\inst{1}
\and M. \, Pohl\inst{2}
\and N. \, Bissantz\inst{3}
           }

\offprints{P. Englmaier}

\institute{
Institut f\"ur Theoretische Physik, Universit\"at Z\"urich, Winterthurerstr. 190, 8057 Z\"urich, Switzerland;
\email{Peter.Englmaier@physik.uzh.ch}
\and
Department of Physics and Astronomy, Iowa State University, Ames, IA 50011-3160, USA;
\email{mkp@iastate.edu}
\and
Fakult\"at f\"ur Mathematik, Ruhr-Universit\"at Bochum, 44780 Bochum, Germany;
\email{Nicolai.Bissantz@ruhr-uni-bochum.de}
}

\authorrunning{Englmaier, P.}
\titlerunning{The Milky Way Spiral Arm Pattern}

\abstract{
A complete map of the 3D distribution of molecular (CO) gas was constructed using a realistic dynamical model of the gas flow in the barred potential of the Milky Way. The map shows two prominent spiral arms starting at the bar ends connecting smoothly to the 4-armed spiral pattern observed in the atomic hydrogen gas in the outer Galaxy. Unlike previous attempts, our new map uncovers the gas distribution in the bar region of the Galaxy and the far side of the disk. For the first time, we can follow spiral arms in gas as they pass behind the galactic centre.

}
\maketitle{}

\section{Introduction}

\begin{figure}[]
\resizebox{\hsize}{!}{\includegraphics[clip=true,angle=90]{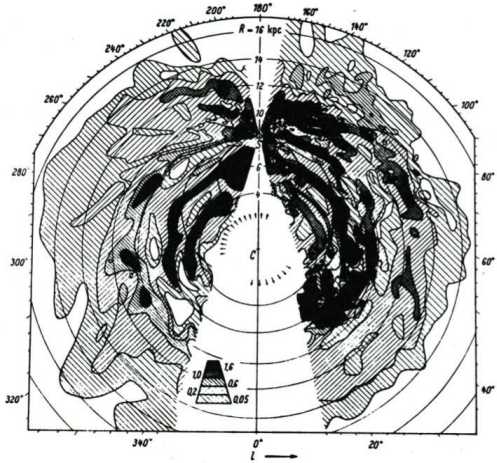}}
\caption{
Map of neutral atomic hydrogen (21-cm line) published by \protect{\cite{Oort58}}; figure taken from the text book  \cite{ElsaesserBook}. The Sun is in the upper part of the plot at $8\,\kpc$.
}
\label{oort}
\end{figure}

Half a century ago, \cite{Oort58} used the Leiden/Sydney 21 cm line survey to construct a map of the neutral atomic hydrogen gas distribution for the Milky Way (Fig.~\ref{oort}). This map, the first large scale map of the Milky Way's gas distribution - mostly located in spiral arms, was disturbed by distance errors and excluded the inner part of the Galaxy as well as the region beyond. The apparent expansion of the Galactic centre region was speculated to be due to a bar, and this view has been generally accepted in the last decade. The method used by Oort, however, assumed circular rotation of the gas, since distance information was not known, and therefore was not applicable in the innermost $\sim5\,\kpc$'s of the Galaxy. The map showed spiral arms, but also many fingers pointing towards earth (the ``Finger-of-God'' effect), and the pattern was incomplete in direction of the centre and anti-centre. There was also an apparent difference between spiral arms on the left and right side of the Galaxy; not a single spiral pattern could fit the observations.

Since then, many studies attempted to chart the spiral arm pattern of the Milky Way in several tracers, but often assuming circular rotation laws for translating radial velocity into distance. Some tracers follow a 4-armed spiral pattern, while others only follow 2 arms (see \cite{Vallee95} for a review). The perhaps most successful attempt to chart spiral arms, was achieved by \cite{GG76}, who used HII regions and also a circular rotation law for distance estimation, or more direct methods for nearby objects.

More recently, \cite{NakanishiSofue} used the $^{12}$CO ($J=1-0$) survey data of \cite{Dame01} to recover the 3D distribution of the molecular gas in the Milky Way. Again, a circular rotation law was assumed, and the area beyond the Galactic centre was excluded. The face-on view is compatible with a 4-armed spiral pattern.

In the outer disk, spiral arms have been traced by analyzing the  HI layer thickness \citep{Levine06}, again finding at least four spiral arms.

\section{Method}

\cite{Pohl08} used the velocity field from the standard model of \cite{Bissantz03} to recover the gas distribution in the Milky Way using a probabilistic method to match the observed CO gas distribution from \cite{Dame01} to the model prediction along the line-of-sight. The underlying kinematic model is not a simple circular rotation law, but has been calculated from a realistic mass model including a tri-axial model of the bulge/bar which has been determined using the observed COBE/DIRBE near-IR light distribution \citep{BissantzGerhard02}. At radii larger than $7\,\kpc$ we use a circular rotation law (after a smooth transition).

\begin{figure*}[]
\resizebox{\hsize}{!}{\includegraphics[clip=true]{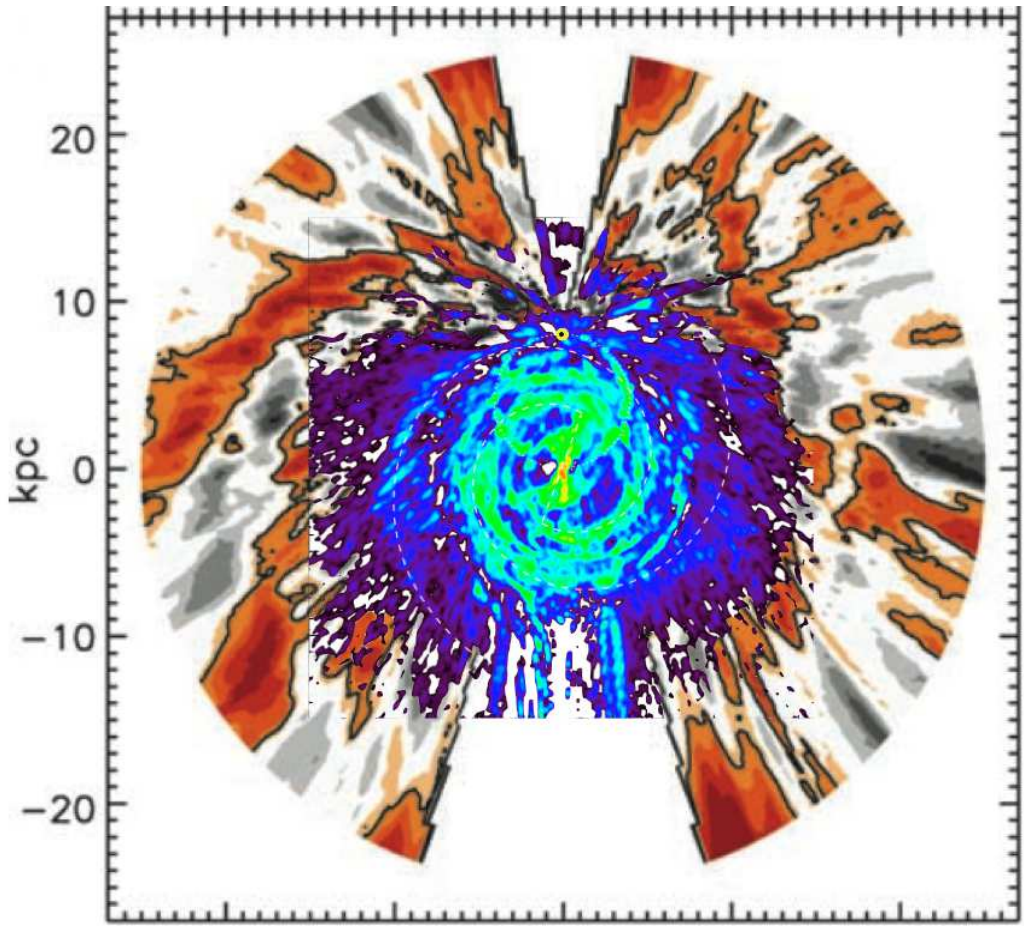}}
\caption{
Map of 3D molecular gas distribution in the inner galaxy (inner part) from \cite{Pohl08}, and thickness HI layer in the outer disk from \cite{Levine06}. The Sun is indicated by the yellow dot.
}
\label{gas}
\end{figure*}

When multiple distances are permitted by the model for a given measured signal, the signal is distributed over the allowed distance bins according to certain weights. These weights have been chosen to avoid placing gas at unrealistic large distances or above or below the warping and flaring plane. Our approach is based on the ideas of regularization methods which are
used commonly e.g. for non-parametric reconstruction problems. Comparison with a mock density model allows us to identify artefacts caused by the inversion. The resulting map for the gas distribution is shown in Fig.~\ref{gas} (blue-green inner part) together with the HI layer thickness (red-gray outer part) from \cite{Levine06}. Major inversion artefacts in this map are: the circle between Sun and galactic centre, the linear structure behind the galactic centre on along the line-of-sight, and the structure seen beyond the solar radius in the far side of the disk.

For further details of the method we refer the reader to \cite{Pohl08}.

\section{Interpretation}

\begin{figure*}[]
\resizebox{\hsize}{!}{\includegraphics[clip=true]{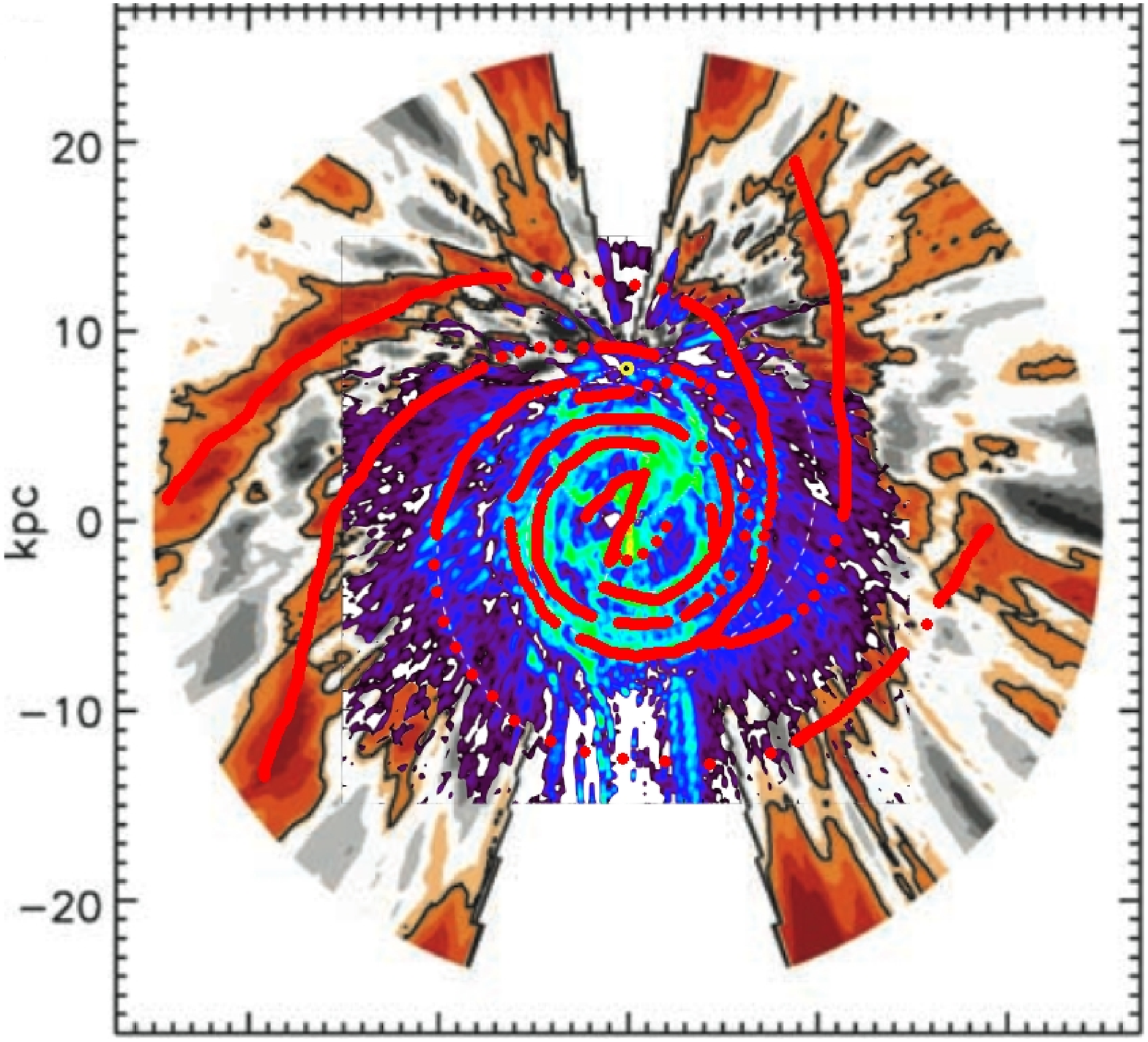}}
\caption{
The inner Galaxy is dominated by the bar and a symmetric 2-armed spiral pattern plus the 3-kpc-arms ending at $\sim3\,\kpc$. The outer galaxy is more irregular with about 4 spiral arms.
}
\label{arms}
\end{figure*}

\subsection{Two or four spiral arms?}

When we trace by eye the spiral arms in Fig.~\ref{gas} starting at the bar ends, the situation becomes complicated when we reach $\sim7\,\kpc$ in radius. Spiral arms seem to end or branch and any picture drawn is highly subjective. However, we can trace the arms with confidence at small and large radii. On the other hand, we can make a sensible connection between the spiral arms, since arms cannot cross, only branch. When we interpret the spiral pattern this way, we can draw the pattern shown in Fig.~\ref{arms}, a 2-armed spiral pattern in the inner Galaxy, which branches in two more arms at about the solar radius. Similarly, there seems to be some indication of short branches starting of the minor axis of the bar when the spiral arms pass by near the Langrangian points of the bar. Those short branches might be due to the assumed kinematical gas flow model. Unfortunately, the 3-kpc-arms are only hinted at in our map. This is a consequence of the poor fit of the gas dynamics near the 3-kpc-arm, which causes the gas from the near 3-kpc-arm to be broken into two pieces. Nevertheless it can be identified in the map and we also see a weak signal for the counter arm.

Since the spiral arms here are matched to structure seen in the deprojected gas distribution map, there is no reason to expect symmetry in the derived spiral pattern. However, surprisingly we find an almost perfect 180-degree rotational symmetry in the inner Galaxy. Since artefacts, real spurs, and gaps in the map are not expected to be symmetrically distributed, we conclude that the observed symmetry and the 2-armed spiral pattern must be real.

In the transition region, at 7..8 kpc galactic radius, we observe spiral arm branching which seems to occur at two locations which are not 180 degree apart in azimuth. When we overlay the spiral pattern with the pattern rotated by 180 degrees, we observe that the spiral arms from both patterns alternately cross and interleave each other. This seems to indicate, that the outer galaxy spiral pattern is a superposition of even and odd spiral modes. The Galaxy nevertheless appears rather symmetric 4-armed, but this might be an illusion. It remains to be seen, if the situation for the Milky Way is similar to the hidden 3-armed spiral mode found in many late type spirals as observed by \cite{Elmegreen92}. If true, three arms should be closer together, while one stronger arm, resulting from two superimposed arms,  should be more isolated. 

\begin{figure*}[]
\resizebox{0.31\hsize}{!}{\includegraphics[clip=true]{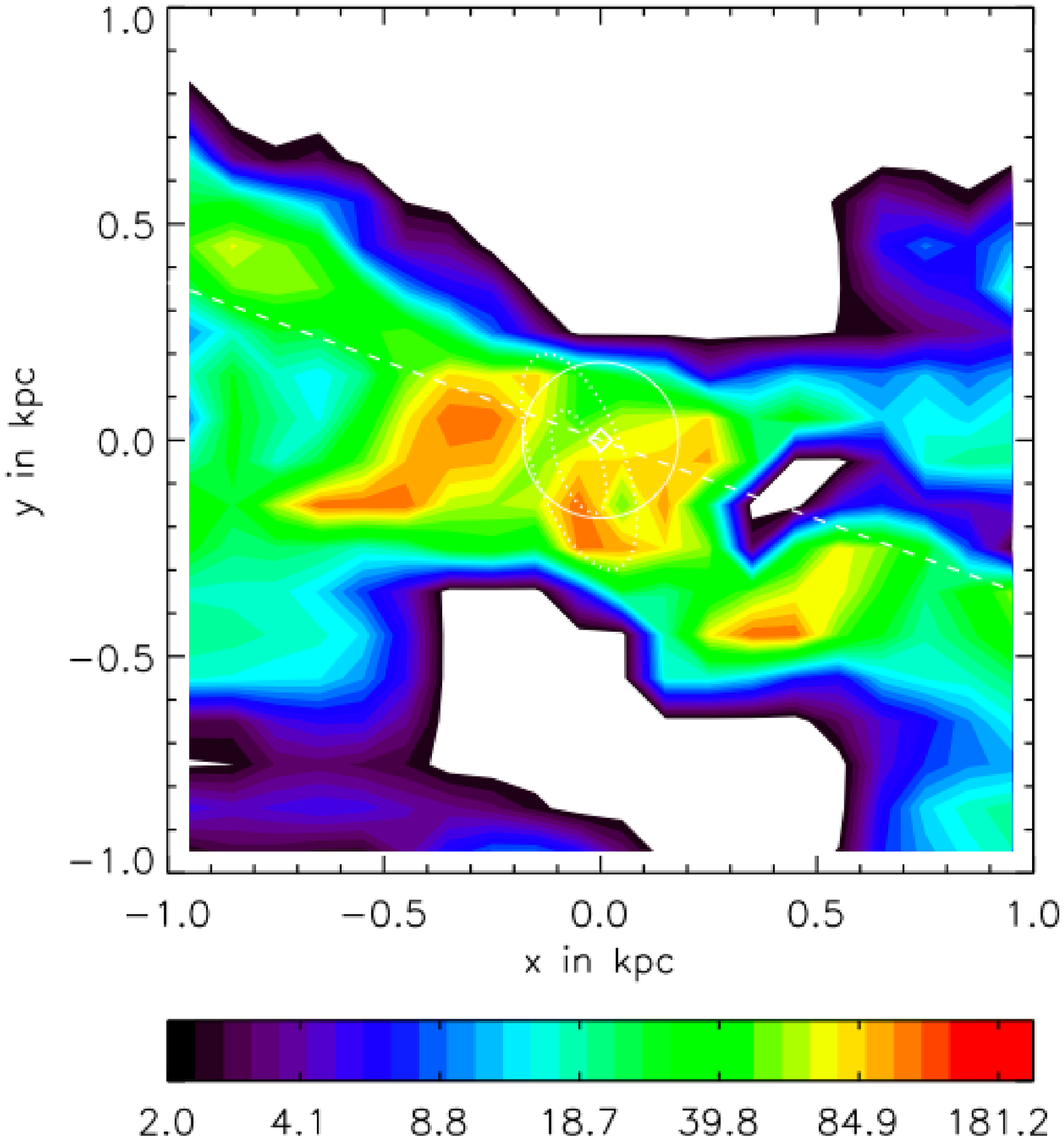}}
\quad\quad
\resizebox{0.5\hsize}{!}{\includegraphics[clip=true,angle=90]{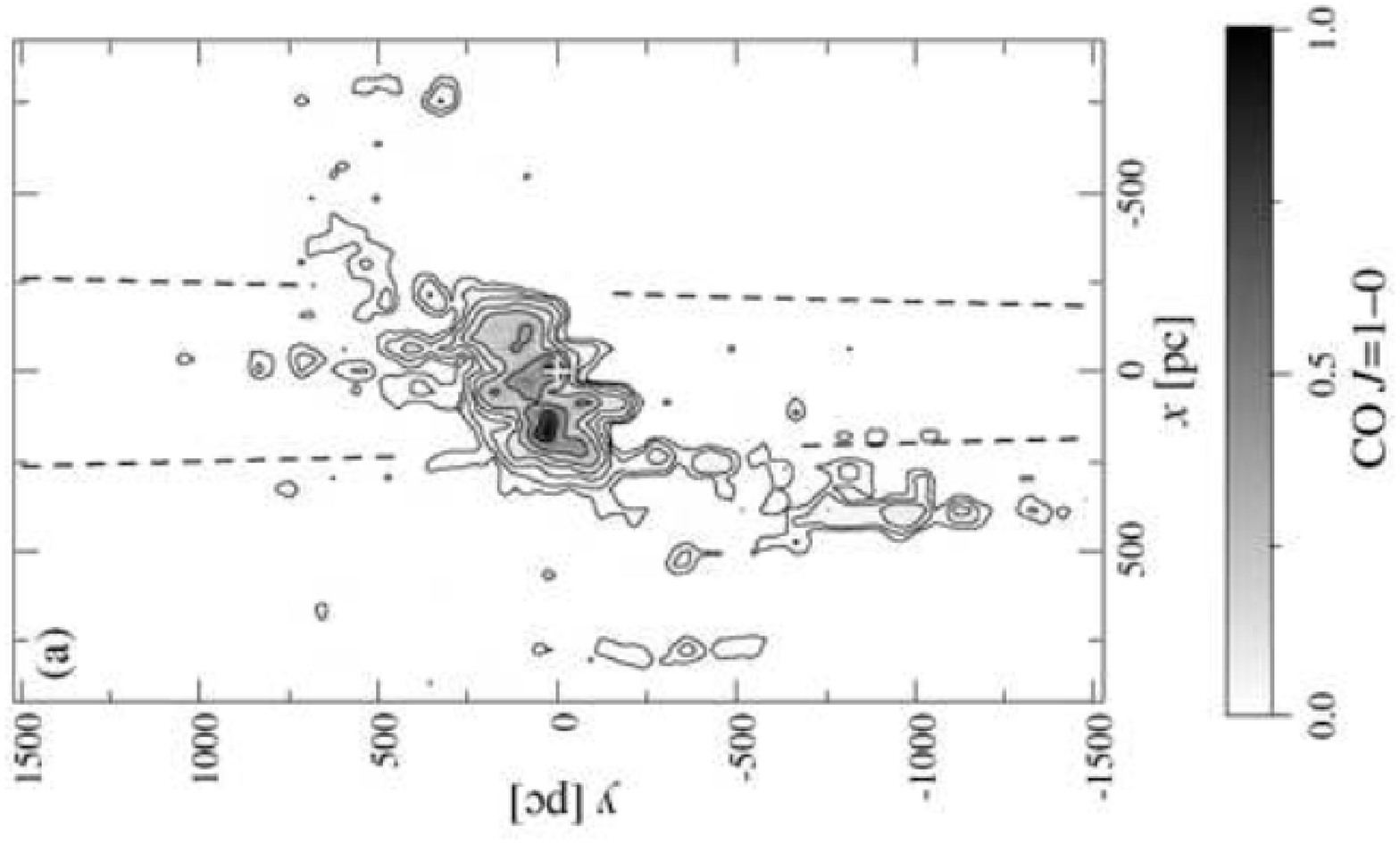}}
\caption{
The centre of the galaxy. {\bf{Left}}: \cite{Pohl08}, and {\bf{right:}} \cite{Sawada06}. The Sun is located on the right at $8\,\kpc$. Horizontal features are indicative of distance errors. The stellar bar in our model is oriented along the dashed line.
}
\label{centre}
\end{figure*}

\subsection{The 3-kpc-arms}

The 3-kpc-arm is a well known and studied feature of the (l,v)-diagram. Its main characteristics are, that it passes in front of the galactic centre with a large radial velocity of $53\,\kms$ which indicates that it is driven by the bar into non-circular motion. 
Alternatively, the 3-kpc-arm's peculiar non-circular motion and apparent lack of a counter-arm, led \cite{Fux99} to the conclusion, that the arm is pushed around by a $m=1$ mode in the inner disk, caused by an off-centre bar tumbling around the centre with a low pattern speed of $\sim20..30\,\kmskpc$. The counter arm is pushed to much larger non-circular velocities explaining the $+135\,\kms$ feature.

Aligned with the 3-kpc arm is a group of OH/IR stars, which has lead to the interpretation of a material arm \citep{Sevenster99} in the context of the \cite{Fux99} model, but can also be understood in terms of the OH/IR progenitors being formed near the Lagrange points at corotation and perpendicular to the bar \citep{Englmaier00}. 

Very recently \cite{Dame08} have found that a possible complementary far 3-kpc-arm has been overlooked in the $(l,v)$-diagram, which is sursprisingly symmetric to the near 3-kpc-arm. While \cite{Dame08} have so far only uncovered part of the structure, it seems like a long searched for piece of the galactic puzzle has fallen in place. The vertical thickness of the new arm is about half the value for the near arm, which places it at the same distance from the galactic centre as the near arm. Moreover, the two arms, if symmetric, allow estimation of the position angle of the bar. By assuming that the arms are bisymmetric and start on the major axis of the bar, we can estimate that the bar's position angle is in the range of 20 to 40 degrees. The value depends critically on the longitude extent of the far arm. \cite{Dame08} find the arm extends to $l=-7..-8\deg$ (corresponding to $25..20\deg$ for the bar angle), or $l=-12\deg$ if two isolated clouds which lie in the continuation of the observed part of the far 3-kpc-arm also belong to it (Dame, priv. comm.). This, however, is unlikely, because the far arm would then not appear as a coherent structure, very unlike the near arm.

\subsection{Inner Galaxy}

In the central kpc of the reconstructed map, we observe a ring with radius $\sim200\,\pc$, which is off-centre, density peaks before and behind the galactic centre, and gas along the leading edges of the nuclear bar (see Fig.~\ref{centre}; the dashed line indicates the position of the bar). We can make a direct comparison with the results from \cite{Sawada06}, which used a non-kinematical method to map the molecular gas in the same region.
 \cite{Sawada06} found a slightly different distribution.
Our reconstruction of the inner Galaxy is resolution limited, smoothing out the innermost few $100\,\pc$.  Both methods have distance errors, causing ``Finger-of-God'' structure pointing to earth. 
  
  In \cite{Sawada06}, the clump named Bania's clump 2 \citep{Bania77} is stretched along the line-of-sight. A similar structure exists in our map, but stretched out over a smaller distance (at (x,y)=(0.4,-0.4) kpc). Another clump on the far side, but somewhat closer to the centre, is seen in our map (at (x,y)=(-0.3,0.1), and might be the counter object to Bania's clump 2. Or, it might be misplaced in distance and it should truly sit at (x,y)=(0,0.1).

\subsection{Outer Galaxy}

In the outer galaxy we can compare and continue our map with the map produced by \cite{Levine06}. 
They used the thickness of the HI layer to trace spiral arms in the outer galaxy out to more than 20 kpc in radius. The maps overlap at $r\sim8\,\kpc$ allowing us to compare and continue the spiral arms between the two studies. We find both maps to agree very well, all four arms can be continued into the map provided by \cite{Levine06}.

\section{Comparison with other studies}

In the previous section, we already compared to \cite{Sawada06} in the inner galaxy and \cite{Levine06} in the outer galaxy. Another study, which used the same data and created a face-on map of the Milky Way was recently done by \cite{NakanishiSofue}. We find excellent agreement in the outer part of the model and can cross-identify features in both studies.

The main difference seems to be the interpretation in terms of spiral arms. 

\section{Conclusions}

The Milky Way has four symmetric spiral arms in the inner part, two of which, the near and far 3-kpc-arm, end inside corotation, two other arms start at $\sim4\,\kpc$ on the major axis of the bar, continue through corotation, and branch at $\sim7\,\kpc$ into four spiral arms which continue to  $\sim20\,\kpc$.

The outer pattern very likely is a superposition of $m=2$ and $m=3$ spiral density waves, as has been observed in external galaxies similar to the Milky Way.

The observed symmetries might be useful in future attempts to invert the gas distribution. One of the spiral arm branches occurs near to us and might leave an imprint in the local stellar velocity distribution. Models of the gas dynamics should take the symmetry of the two 3-kpc-arms into account.


\bibliographystyle{aa}

\end{document}